\documentclass[a4paper,11pt]{article}
\pdfoutput=1 % if your are submitting a pdflatex (i.e. if you have
\usepackage{jinstpub} % for details on the use of the package, please
\usepackage[version=3]{mhchem}
\usepackage{siunitx}
\usepackage{natbib}
\usepackage{amsmath} % or simply amstex

\usepackage{verbatim}
\usepackage{multirow}

\usepackage[english]{babel}
\usepackage{hyperref}
\addto\extrasenglish{
    
}

\usepackage{setspace}
\usepackage{graphicx}
\usepackage{amsmath}
\usepackage{siunitx}

\usepackage{pbox}
\usepackage{xcolor}

\newcolumntype{"}{@{\hskip\tabcolsep\vrule width 1pt\hskip\tabcolsep}}
\DeclareSIUnit[number-unit-product = \;]\year{yr}
\DeclareSIUnit\parsec{pc}
\DeclareSIUnit\torr{Torr}
\DeclareSIUnit\centimeter{\centi \meter}
\DeclareSIUnit\sq{\ensuremath{\Box}}
\usepackage{textcomp}
\usepackage{caption}
\usepackage{subcaption}
\usepackage{url}
\usepackage{ctable}

\usepackage{lineno}
\title{High Negative Ion Gain MMThGEM-Micromegas Detector for Directional Dark Matter Searches}

\author[a,b,c,d,1]{A.G. McLean,\note{Corresponding author.}}
\author[e]{S. Higashino,}
\author[b,d]{R.R. Marcelo Gregorio,}
\author[e]{K. Miuchi,}
\author[d]{and N.J.C. Spooner }

\affiliation[a]{School of Physics, Chemistry and Earth Sciences, Adelaide University, North Terrace Campus, Adelaide, SA 5005, Australia}
\affiliation[b]{Department of Nuclear Physics and Accelerator Applications, Australian National University, Garran Road, ACT 2601, Canberra, Australia}
\affiliation[c]{ARC Centre of Excellence for Dark Matter Particle Physics, Australia}
\affiliation[d]{Department of Physics and Astronomy, University of Sheffield, South Yorkshire, S3 7RH, United Kingdom}
\affiliation[e]{Department of Physics, Kobe University, Rokkodaicho, Nada Hyogo 657-8501, Japan}

\emailAdd{alasdair.mclean@adelaide.edu.au}

\abstract{Low pressure gaseous Negative Ion Time Projection Chambers (NITPCs) have been used previously by the DRIFT experiment to search for a directional Dark Matter (DM) signature. The main challenge with using a Negative Ion Drift (NID) gas target is the significantly lower gas gains to which they are typically limited. Recently, a MMThGEM device has been successfully demonstrated as an excellent gain stage device in the NID gas SF$_6$; capable of producing gas gains comparable with the electron drift gas CF$_4$. The next major challenge is to extend this high gain capability to multi-dimensional readout for the purpose of particle track reconstruction. The MMThGEM is therefore ideal for coupling to a strip readout detector like a Micromegas to achieve a high gain multi-dimensional Negative Ion (NI) readout plane, which is potentially suitable for the scale up required by future searches proposed by the CYGNUS consortium. In this paper, the first high gain demonstration of such a MMThGEM-Micromegas detector in low pressure SF$_6$ is described. This includes detector characterisation in a small test vessel resulting in the largest NI gas gain ever reported, 1.22 $\pm$ 0.08 $\times$ 10$^5$, and directionality with alpha particles. Finally, this gain characterisation and tracking capability is leveraged to measure the energy and range of events, and identify those consistent with Nuclear Recoils (NRs), in a large cubic metre scale volume of SF$_6$ for the first time.}


\begin{document}
\maketitle
\flushbottom
\raggedbottom

\section{Introduction}
\label{sec:intro}

%The direct search for Weakly Interacting Massive Particles (WIMPs), which could account for the 85\% of the total mass in the Universe known as Dark Matter (DM), has now reached the limitations of the neutrino fog \cite{solarNsPandaX,solarNsXENON, solarNsLZ}. The neutrino fog is now set to impact the discovery potential of WIMPs in leading detector searches like LZ, XENONnT, PandaX-4T and future efforts of the XLZD collaboration. Without a precise understanding of neutrino fluxes on Earth, these experiments will likely struggle to provide conclusive evidence for the direct detection of such a particle. A strong alternative signature for the positive identification of WIMP-like DM could be sought through directional detection \cite{OHare2021}. 

The direct search for Weakly Interacting Massive Particles (WIMPs), potential candidates for the 85\% of the Universe’s mass attributed to Dark Matter (DM), has reached the sensitivity limit imposed by the neutrino fog \cite{solarNsPandaX,solarNsXENON, solarNsLZ}. This background now constrains the discovery potential of leading experiments such as LZ, XENONnT, PandaX-4T, and future XLZD detectors. Without precise knowledge of terrestrial neutrino fluxes, these experiments may struggle to deliver conclusive WIMP detection. Directional detection offers a promising alternative for the unambiguous identification of WIMP-like DM \cite{OHare2021}.

Such a directional search would allow for discrimination between neutrinos, which largely originate from the Sun; and WIMPs, which appear to originate from the Cygnus constellation due to the motion of the Solar System through the Milky Way Galaxy. As these origins are separable on Earth at all times, this approach offers the potential for an unimpeded search below the neutrino fog \cite{Vahsen2021}. Furthermore, the galactic origins of this signal would be unambiguous due to its angular modulation caused by the rotation of the Earth on its own axis \cite{Morgan2003}. In contrast to measurements of annually modulated event rates, this angular modulation could not be conflated with terrestrial sources \cite{Klinger2015,Kudryavtsev2010,Davis2014}. A directional measurement is therefore considered to be the most conclusive method for WIMP discovery.

Current leading two-phase detector technologies have been found incapable of directional detection \cite{DarkSide-20k}, therefore an alternative technology is required for this kind of search. Low pressure gaseous Time Projection Chamber (TPC) methods have been used previously, by experiments like DRIFT \cite{Bat2016} and NEWAGE \cite{Shimada2023}, to search for directional Nuclear Recoils (NRs) resulting from WIMP interactions. The low pressure allows for longer tracks of ionisation in the gas to better match the spatial resolution of readout technologies. Further to this, the DRIFT experiment conducted a directional WIMP search using the Negative Ion Drift (NID) gas CS$_2$ for improved reconstruction of events compared to conventional electron drift gases. The use of SF$_6$ in future searches, like that proposed by the CYGNUS collaboration \cite{Vahsen2020}, is being investigated as an alternative NID gas target. This is because SF$_6$ is safer to work with and offers improvements in Spin-Dependent (SD) cross-sections with potential WIMP candidates \cite{Phan2017}. However, one challenge of using NID gases has always been their limited gas gain, caused by inefficiency of stripping electrons from the NI before charge amplification can occur, resulting in a reduced sensitivity to low energy recoil events.

Recent results with a promising Multi-Mesh Thick Gaseous Electron Multiplier (MMThGEM) device in SF$_6$ have shown that gas gains in a NID target can be comparable to conventional gases, like CF$_4$, provided that significant attention is paid to the design and optimisation of the gain stage device \cite{McLean2024, Amaro2024}. This gain stage device is therefore an ideal candidate for coupling to a strip readout detector like a Micromegas in order to achieve multi-dimensional directionality. In this paper, this coupled detector is introduced, followed by a description of the experimental setup in a test vessel. The effective gas gain is established on the strips and the detector's directional response to alpha particles is evaluated. Finally the response of the detector to NRs is tentatively explored in a large metre cubed scale vessel.

\section{Detector Configuration and Experimental Setup}
\label{sec:Setup}

\begin{comment}
\begin{figure}[b]
    \centering
    \includegraphics[width=0.7\textwidth]{Images/MMThGEM-MicromegasDiag.png}
    \caption{Cross sectional diagram of the coupled MMThGEM-Micromegas detector depicted
    with a 3 cm drift length used during small scale testing.}
    \label{fig:detector_config}
\end{figure}
\end{comment}

\begin{figure}[b]
    \centering
    \includegraphics[width=0.9\textwidth]{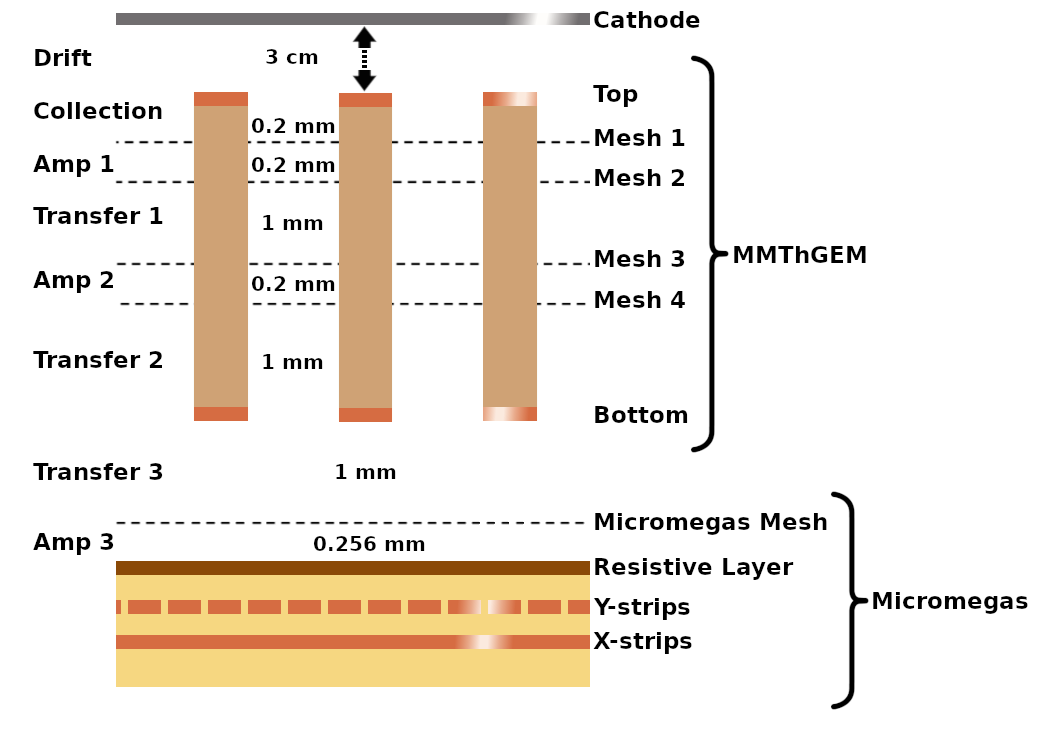}

    \captionsetup{justification=centering}
    \caption{Cross sectional diagram of the coupled MMThGEM-Micromegas detector.}
    \label{fig:detector_config}
\end{figure}

The TPC used in this work is depicted in \autoref{fig:detector_config} and consists of a cathode, a MMThGEM and a Micromegas. A MMThGEM is a MicroPattern Gaseous Detector (MPGD) similar in structure to a Thick-Gaseous Electron Multiplier (ThGEM) \cite{Bressler2023} which incorporates additional mesh layers \cite{Olivera2018}. The two stage MMThGEM device, used in this work, consists of top and bottom electrode layers and 4 intermediate mesh layers, for improved amplification properties. These electrode layers establish a collection field, and pairs of amplification and transfer fields; lower field strengths are utilised in the collection and transfer fields to facilitate the transfer of NIs while larger fields utilised in the amplification fields are designed to initiate electron stripping and avalanche. This design has recently demonstrated significant sub-10$^5$ gas gains in SF$_6$ \cite{McLean2024} and has therefore been identified as a suitable gain stage device for coupling with a multi-dimensional readout plane like a Micromegas.

A Micromegas detector is a type of MPGD which utilises a parallel plate avalanche region, established between a mesh electrode and a micro-strip plane \cite{Attie2021}. The Micromegas used in this work was mounted 1 mm below the MMThGEM and consists of a Diamond Like Carbon (DLC) resistive anode layer on top of orthogonal x and y micro-strip electrode planes. The y-strips are situated above the x-strips and have a width of 100 µm and 220 µm respectively with a pitch of 250 µm. The Micromegas has an active area of 10 $\times$ 10 cm$^2$, identical to the MMThGEM.

\begin{table}[b]
    \centering
    \caption{Electric field strengths used under configuration A (optimised high gain MMThGEM voltage settings \cite{McLean2024}) and configuration B (lower gain voltage settings to avoid sparking during alpha measurements).}
    \label{tab:field_configurations}
    \begin{tabular}{lcc}
    \hline
    \textbf{Detector Field} & \multicolumn{2}{c}{\textbf{Field Strength (V/cm)}} \\
    \cline{2-3}
                             & \textbf{Configuration A} & \textbf{Configuration B} \\
    \hline
    Drift            & 333 & 333 \\
    Collection          & 2000 & 1835 \\
    Amplification 1 \& 2     & 28000 & 26615 \\
    Transfer 1 \& 2         & 900 & 826 \\
    Transfer 3     & 360 & 333 \\
    Amplification 3    & 20703  & 19531 \\
    \hline
    \end{tabular}
    \end{table}

For detection to occur in this coupled detector, the charge from ionising events in the drift region is first drifted towards the top of the MMThGEM and then funneled into the holes of the MMThGEM under the influence of the collection field. This charge is then accelerated by the large field strength of the first amplification field resulting in a cascade of ionisation which serves to amplify the charge. This amplified charge is transferred to the second amplification field, by a smaller transfer field strength, where it is accelerated and amplified for a second time. The charge is then transferred towards the Micromegas by the second and third transfer fields before being amplified again by the third and final amplification field in the Micromegas. Finally, the amplified charge is measured on 32 Micromegas y-strips instrumented with LTARS2018 charge sensitive electronics \cite{Kosaka2022}. This amounts to a total instrumented area of 7.85 mm $\times$ 10 cm.

The MMThGEM biasing was handled by two HV supplies and a resistor chain solderd to the biasing connections. These resistor values were derived from a previous optimisation in 40 Torr SF$_6$ \cite{McLean2024}. Two additional HV supplies were used for biasing the cathode and Micromegas mesh separately, while the resistive layer and strips of the Micromegas were connected to ground. All four HV channels were provided by CAEN A7030DP and A7030DN modules. \autoref{tab:field_configurations} presents two operating configurations that were demonstrated to be stable (i.e., no electrical sparking was observed): Configuration A, employing the optimised high gain settings, and Configuration B, lower voltage settings used during alpha measurements.

The charge calibration of the electronics was performed by injecting square waves via a 1pF capacitor into each channel. The amplitude was varied from 50 - 100 mV in increments of 10 mV and then from 100 - 2400 mV in increments of 100 mV, due to the dynamic range of the LTARS2018 electronics. A signal integral calibration was used to determine the charge due to the slow rise time of NI signals. The charge calibration was applied globally because the variation across all 32 channels was found to be small; for example, the pulse height response of the high gain channels measured 10.64 $\pm$ 0.05 mV/fC.

%To begin the detector characterisation, the coupled MMThGEM-Micromegas TPC was first assembled and mounted in the centre of an aluminium plate, shown in \autoref{fig:MMThGEM-MicromegasImage}. This plate was mounted to the door of a small test vessel. Once assembled, the detector was then sealed inside the test vessel by mounting the door.

\begin{figure}[h]
    \begin{subfigure}[b]{0.495\textwidth}
        \centering
        \includegraphics[trim={0.38cm 1.2cm 0.3cm 1cm}, clip, angle=270,width=\textwidth]{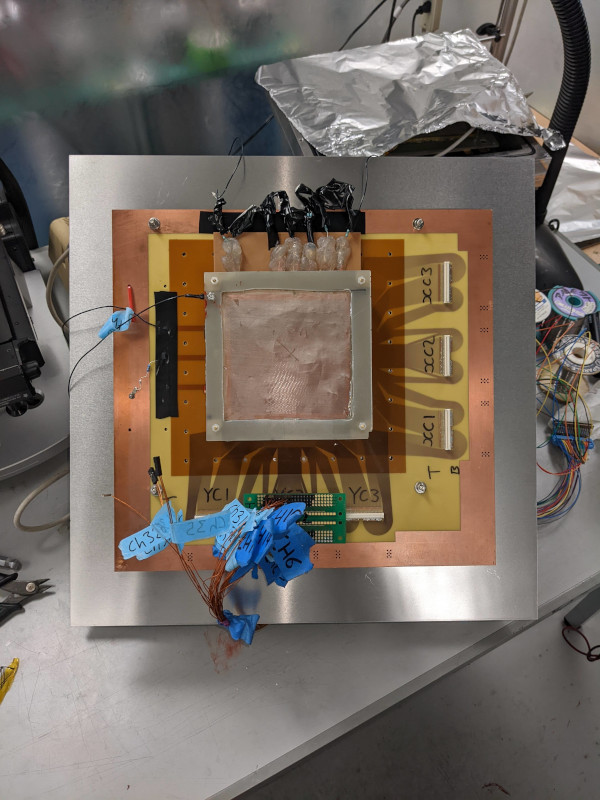}

        \caption{}
        \label{fig:MMThGEM-MicromegasImage}
    \end{subfigure}
    \hfill
    \begin{subfigure}[b]{0.495\textwidth}
        \centering
        \includegraphics[trim={0 0 0 1.885cm},clip,width=\textwidth]{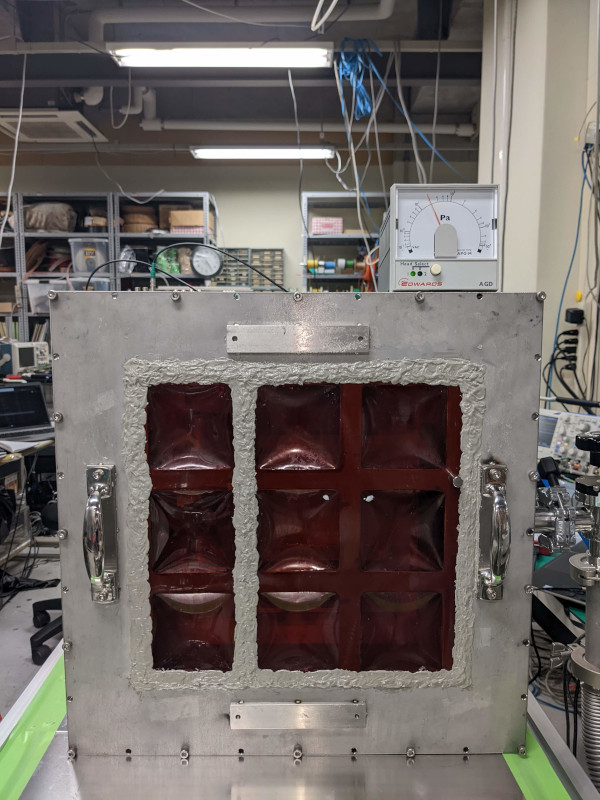}
        \caption{}
        \label{fig:TestVesselKapton}
    \end{subfigure}

    %\includegraphics[trim={8cm 25cm 7cm 22cm},clip, width=0.6\textwidth]{MMThGEM-MicromegasImage.jpg}
    %\caption{test caption }
    \label{fig:test_vessel_images}
    \caption{(a) Image of the coupled MMThGEM-Micromegas TPC assembly mounted to the door of the test vessel. (b) Image of the test vessel with brown kapton window containing the coupled MMThGEM-Micromegas TPC assembly mounted inside.}
\end{figure}

\begin{figure}[h!]
    \centering
    \includegraphics[trim={0cm 0cm 0cm 0cm},clip, width=0.99\textwidth]{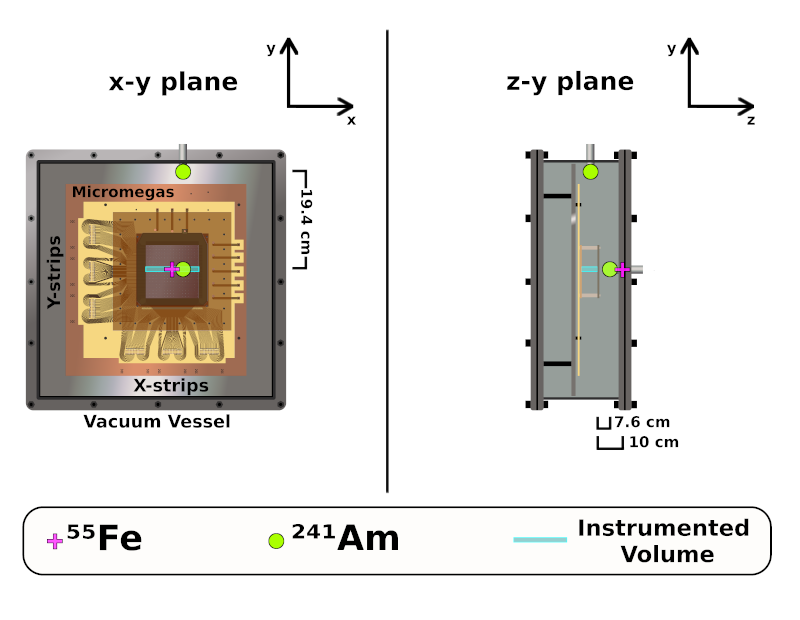}
    \caption{Diagram of small test vessel showing the positioning of radioactive sources around the TPC volume.}\label{fig:Test_vessel_Source_pos}
\end{figure}

For detector characterisation, the MMThGEM-Micromegas TPC was assembled and mounted at the centre of an aluminium plate (\autoref{fig:MMThGEM-MicromegasImage}), which was attached to the door of a small test vessel. The detector assembly was then sealed inside the vessel which also featured a thin kapton window, shown in \autoref{fig:TestVesselKapton}, to aid exposure to radioactive sources.
The test vessel was evacuated and filled with 40 Torr of SF$_6$ and sources of ionising radiation were positioned around the instrumented TPC volume. The sources included an $^{55}$Fe X-ray source (\autoref{sec:GasGain}), 0.8 MBq, and an $^{241}$Am alpha particle source (\autoref{sec:Directional}), $\mathcal{O}$(10 kBq). A diagram was created to illustrate the source positions relative to the instrumented TPC volume and can be seen in \autoref{fig:Test_vessel_Source_pos}.

The panel on the left of \autoref{fig:Test_vessel_Source_pos} shows a cross section of the test vessel in the x-y plane, while the panel on the right shows the z-y plane. The cross section of the instrumented TPC volume can be seen highlighted in cyan above the Micromegas plane. The source positions can be seen indicated by a magenta plus and green circle icons for the $^{55}$Fe and $^{241}$Am positions respectively. 
During the X-ray exposure, the $^{55}$Fe source was positioned externally in the kapton window of the test vessel with an offset in the z-dimension of 10 cm from the instrumented volume. For the purpose of orthogonal directional measurements, the $^{241}$Am source was given a y-axis and z-axis exposure position. During the y-axis exposure, the $^{241}$Am source was positioned internally with a 19.4 cm distance in the y-dimension from the instrumented volume. During the z-axis exposures, it can be seen that the $^{241}$Am source was positioned directly in front of the instrumented volume with a separation in the z-axis of 7.6 cm.

\section{Effective Detector Gas Gain with $^{55}$Fe X-rays}
\label{sec:GasGain}

%In this section, work which evaluates the gas gain measured on the detector strips using the $^{55}$Fe source is presented. The HV biasing was set to be -2900 V, -1900 V, 100 V, and -530 V for the cathode, V$_{in}$, V$_{out}$ and Micromegas mesh respectively. 
%This corresponds to a drift field of 333 V/cm, and conserves the optimised MMThGEM field strengths with a 38.7 V potential drop across the collection field, amplification 1 and 2 field strengths of 28015 V/cm, transfer 1 and 2 field strengths of 870 V/cm. The third transfer and amplification fields provided by the Micromegas have field strengths of 367 V/cm and 20703 V/cm. 

%$^{55}$Fe X-ray events were successfully detected on both the HG and LG channels; a range of voltage settings could not be determined due to the narrow range of stable voltages which conserved the optimised MMThGEM voltage settings and produced signals visible to both HG and LG channels.

\begin{figure}[ht]
    \centering

    \begin{subfigure}{0.5\textwidth}
        \centering
        \includegraphics[trim={0cm 0cm 0cm 1.5cm},clip,width=\linewidth]{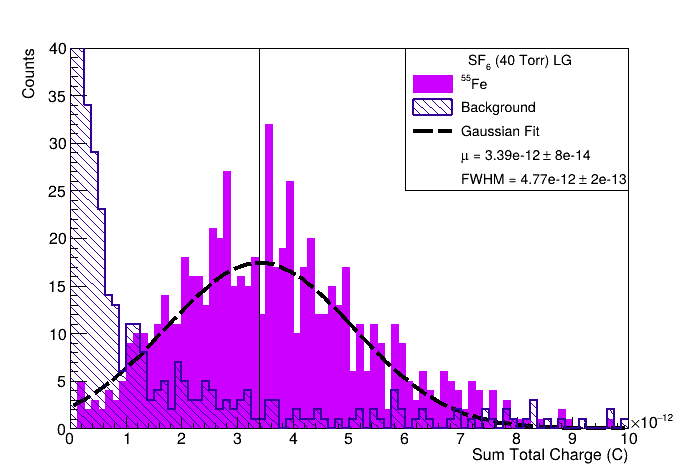}
        \caption{}
        \label{fig:GainKobe}
    \end{subfigure}
    \hfill
    \begin{subfigure}{0.48\textwidth}
        \centering
        \includegraphics[trim={0cm 0.4cm 0cm 0cm},clip,width=\linewidth]{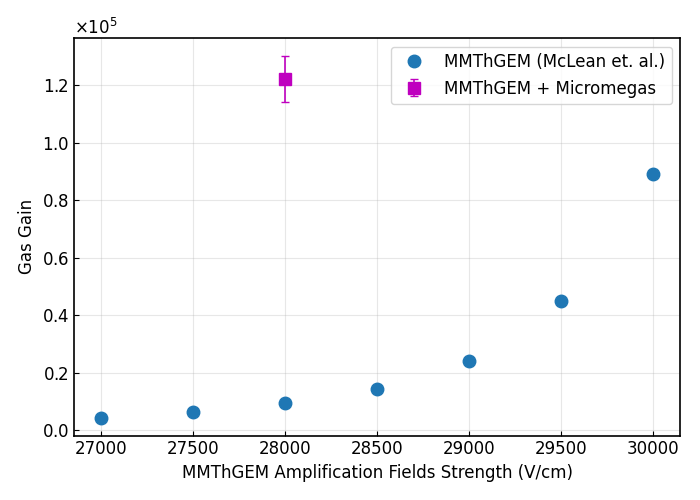}
        \caption{}
        \label{fig:Gain_Curve}
    \end{subfigure}

    \caption{(a) Sum total charge spectrum with the signal integral method as measured on the strips with $^{55}$Fe source (solid magenta) and background (hatched violet) exposures. (b) Gas gain curve from previous MMThGEM optimisation (blue) \cite{McLean2024} and current measurement with the coupled MMThGEM-Micromegas detector (magenta). Error bars which are smaller than the marker size are not observed.}
    \label{fig:CombinedFigure}
\end{figure}

The effective gas gain of the detector is required to determine the electron equivalent energy of events, in this section the gas gain is determined with the $^{55}$Fe source. X-ray events were successfully detected on the Micromegas strips using the high gain Configuration A (see \autoref{tab:field_configurations}). Due to the MMThGEM hole size and charge dissipation in the resistive layer, events span across several strips. Therefore, a strict cut was applied to ensure that the charge was not artificially diminished by falling outside the instrumented area. Events were cut if the central channel number was less than 13 or more than 18. Events passing this cut were binned according to the signal integral, converted to charge via the calibration, and aggregated across all 32 strips. The resulting spectrum is shown in  \autoref{fig:GainKobe}.

%To perform this cut, the channel with the largest amplitude per event was determined, this channel was found to be centred within the cluster, and the  An example of an edge event cut from the gain calculations, with a peak channel number of 27, can be seen on the left in \autoref{fig:EdgeCuts}. An example of a centred event accepted for the gain calculations, with a peak channel number of 13, can be seen on the right of \autoref{fig:EdgeCuts}.

\begin{comment}

\begin{figure}[h!]
    \centering
    \begin{subfigure}[b]{0.49\textwidth}
        \centering
        \includegraphics[trim={0cm 0cm 0cm 0cm},clip,width=\textwidth]{Images/waveHGno98EdgeEventpkch27.png}
        \label{fig:EdgeEvent}
    \end{subfigure}
    \hfill
    \begin{subfigure}[b]{0.49\textwidth}
        \centering
        \includegraphics[trim={0cm 0cm 0cm 0cm},clip,width=\textwidth]{Images/waveHGno427CenterEventPkCh13.png}
        \label{fig:LTARScenterEv}
    \end{subfigure}
    \caption{Example of edge event which was removed by the edge cut (left). Example of centred event accepted for the gain calculations (right).  Vertical offset of channels used for clarity only.}
    \label{fig:EdgeCuts}
\end{figure}

\end{comment}

The $^{55}$Fe spectrum is shown in solid magenta along with a background exposure in hatched violet. A Gaussian function was fit to the $^{55}$Fe exposure data and can be seen as a dashed black line with the mean of the Gaussian indicated by a vertical black line. The gain was determined via the $^{55}$Fe X-ray energy, 5.89 keV, and W-value of SF$_6$, 34 eV \cite{Lopes1986}. It was found that the detector produced an effective gas gain of 1.22 $\pm$ 0.08 $\times$ 10$^5$ with an energy resolution of 1.41 $\pm$ 0.07. The energy resolution is calculated as the FWHM (Full Width at Half Maximum) of the Gaussian divided by the mean. No attempt is made here to disentangle the electron stripping efficiencies or detachment length as this is beyond the scope of this work.

This result is a significant ancillary benefit of the detector coupling because this is the first time a gas gain on the order of 10$^5$ has been achieved with an NID gas; furthermore, this is significantly larger than typical NID gas gains on the order of 10$^3$ \cite{Phan2017,Baracchini, Miyamoto}. This result is credible considering that the previous MMThGEM optimisation, shown alongside the current result in \autoref{fig:Gain_Curve}, is known to be providing an amplification factor of approximately 10$^4$ under Configuration A voltage settings \cite{McLean2024}. The Micromegas mesh therefore contributes an additional amplification factor of approximately 12.2. This large 10$^5$ gas gain will benefit full 3D reconstruction of low energy tracks when the instrumentation is extended to x/y readout, as higher gains have been shown to improve the charge observed on the lower strips \cite{Ghrear}.

\section{Directionality with $^{241}$Am Alpha Particle Tracks}
\label{sec:Directional}

Determining the principal axis of particle tracks and identifying sense via $\frac{dE}{dx}$ signatures is important for directional searches. In this section, work in which a 2-dimentional track reconstruction algorithm was developed, and tested via exposures to an $^{241}$Am alpha particle source, is presented. Due to the highly ionising nature of alpha particles, the field strengths were lowered during these runs to reduce sparking (see Configuration B in \autoref{tab:field_configurations}).

\begin{figure}[h!]
    \begin{subfigure}[]{0.5\textwidth}
        \centering
        \includegraphics[trim={0cm 0cm 0cm 1.8cm},clip,width=\textwidth]{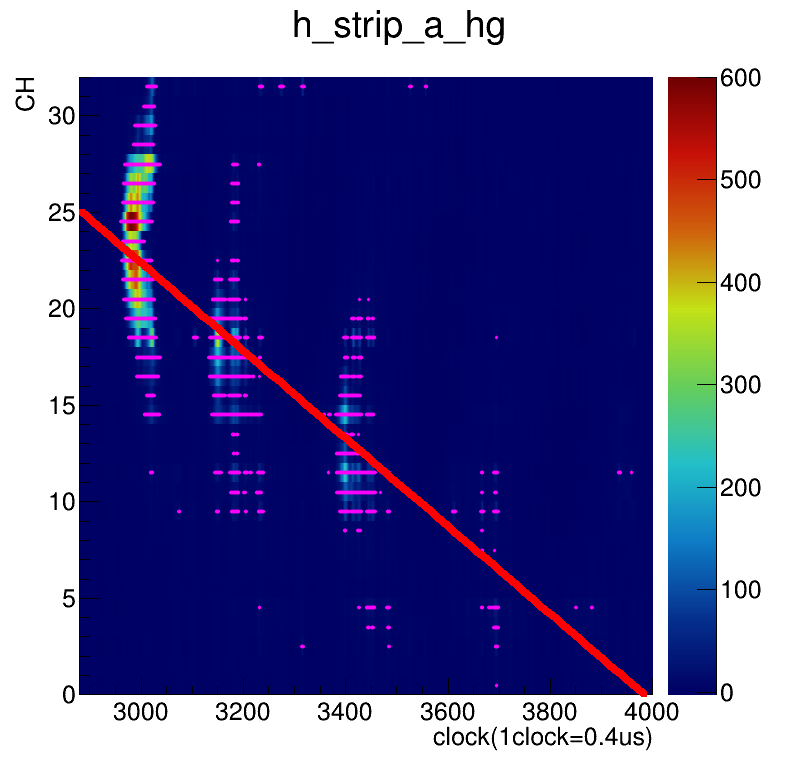}
        \label{fig:TLRz-axis}
    \end{subfigure}
    \begin{subfigure}[]{0.5\textwidth}
        \centering
        \includegraphics[trim={0cm 0cm 0cm 1.8cm},clip,width=\textwidth]{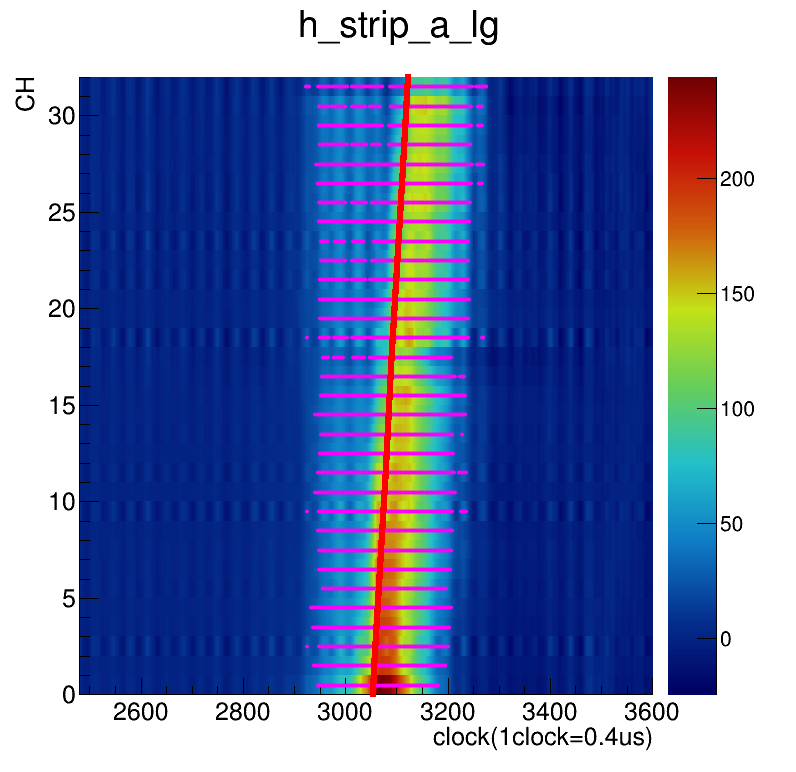}
        \label{fig:TLRy-axis}
    \end{subfigure}
    \caption{An example z-axis exposure event (left) and y-axis exposure event (right). Points above a 40 mV threshold are indicated by magenta markers and the TLR fit is indicated by a red line.}
    \label{fig:TLR_examples}
\end{figure}

Alpha tracks were successfully detected in both the z-axis and y-axis exposure directions, outlined in \autoref{sec:Setup}, examples of which can be seen in \autoref{fig:TLR_examples}. The channel number is plotted against time, where the instantaneous voltage of each channel is indicated by the colour scale in mV, and the points above a 40 mV threshold are indicated by magenta markers. During the z-axis alpha particle exposure, events were observed to have discontinuity between consecutive charge clusters, shown on the left of \autoref{fig:TLR_examples}. This was found to be caused by the MMThGEM hole pitch because the cluster separation was consistent with the 1.2 mm hole pitch.

A Total Linear Regression (TLR), also known as Deming regression, algorithm was developed for determining the principal axis of alpha tracks in the y-axis and z-axis exposures; this accounts for the residuals in both axes and effectively handles the highly rectilinear nature of alpha particle tracks. This algorithm was implemented by first determining all the points above a 40 mV threshold, followed by the conversion of these coordinates to spatial units via the Micromegas strip pitch and ion drift velocity. For each event, the gradient, m, and y-intercept, c, of the track in the z-y plane was calculated as \cite{Glaister2001}:

\begin{equation*}
    m = -p +\sqrt{1+p^2}, \quad c = \bar{Y} - m\bar{Z}, \quad \text{where} \quad p = \frac{S_{ZZ}-S_{YY}}{S_{ZY}},
\end{equation*} $\bar{Y}$ and $\bar{Z}$ are the average y and z coordinates of those above threshold and $S_{ZZ}$, $S_{YY}$, and $S_{ZY}$ are defined as:

\begin{equation}
    S_{ZZ} = \frac{1}{n}\sum_{i=1}^{n} Z_i^2 - \bar{Z}, \quad S_{YY} = \frac{1}{n}\sum_{i=1}^{n} Y_i^2 - \bar{Y}, \quad
    S_{ZY} = \frac{1}{n}\sum_{i=1}^{n} Z_iY_i - \bar{Z}\bar{Y},
\end{equation}where \textit{n} is the total number of points above threshold. By visual inspection, the algorithm performed well for track reconstruction during both perpendicular exposures. The result of this algorithm applied to a z-axis and y-axis exposure event can be seen overlaid in red in \autoref{fig:TLR_examples}.

\begin{figure}[h!]
    \centering
    \includegraphics[trim={0cm 0cm 0cm 0cm},clip,width=0.7\textwidth]{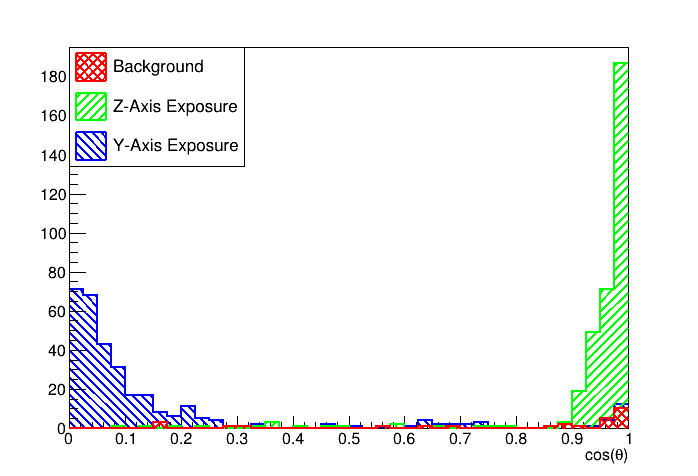}
    \caption{Distribution of the angle between the principal axis of an event and the z-axis determined via the TLR algorithm.}
    \label{fig:AngularSep}
\end{figure}

The TLR algorithm was applied to all events in the z-axis and y-axis exposures and $\theta$, the angle between the z-axis and the reconstructed track, was calculated. The results of which, shown in \autoref{fig:AngularSep}, display clear angular separation between events from the two orthogonal exposures. The z-axis events peak at $\theta$ = 0$^\circ$ while the y-axis events peak at $\theta$ = 90$^\circ$. This result highlights the effectiveness of the TLR algorithm for accurately determining the principal axis in 2-dimensions. Further to the reconstruction of the principal axis, the observed $\frac{dE}{dx}$ signatures of the alpha tracks in \autoref{fig:TLR_examples} are qualitatively consistent with the expected bragg peak direction. This result is significant because it demonstrates that charge asymmetries, and therefore directional sense, can be resolved with this detector.

\begin{comment}
\begin{figure}[h!]
    \begin{subfigure}[]{0.5\textwidth}
        \centering
        \includegraphics[trim={1cm 0cm 1cm 1cm},clip,width=\textwidth]{Images/Alpha_Track_Recon_cosTheta_combined_alt_legend.png}
        \label{fig:AngularSep}
    \end{subfigure}
    \begin{subfigure}[]{0.5\textwidth}
        \centering
        \includegraphics[trim={0cm 0cm 1cm 0.7cm},clip,width=\textwidth]{Images/alphaRangesSRIM_40TorrSF6.png}
        \label{fig:SRIMalpha}
    \end{subfigure}
    \caption{Distribution of the angle between the principle axis of an event and the z-axis determined via the TLR algorithm (left). Bragg curve simulated with 10,000 5.5 MeV alpha particles in 40 Torr of SF$_6$ using SRIM. Hatched areas indicate the section measured by the instrumented strips (right).}
    \label{fig:AlphaResults}
\end{figure}
\end{comment}

\begin{comment}

A SRIM (Stopping and Range of Ions in Matter) \cite{SRIM} simulation of 5.5 MeV alpha particles in 40 Torr of SF$_6$ was performed and the regions of interest for both the z-axis and y-axis exposures can be seen on the right of \autoref{fig:SRIMalpha}, highlighted in green and blue respectively. The intensity of the charge clusters in the z-axis exposure event increases from right to left, in the direction of the alpha particles motion. While the intensity of the measured charge in the y-axis exposure event decreases along the length of the track; the alpha particle is travelling from channel 0 to 31. This result is significant because it demonstrates that charge asymmetries, and therefore directional sense, can be resolved with this detector.

\begin{figure}[h!]
    \centering
    \includegraphics[trim={0cm 0cm 0cm 0cm},clip,width=0.6\textwidth]{Images/alphaRangesSRIM_40TorrSF6.png}
    \caption{Bragg curve simulated with 10,000 5.5 MeV alpha particles in 40 Torr of SF$_6$ using SRIM. Hatched areas indicate the section measured by the instrumented strips. }
    \label{fig:SRIMalpha}
\end{figure}

\end{comment}

%These regions of interest correspond to the separation between the alpha source and instrumented volume during each exposure. As shown, the observed increasing and decreasing $\frac{dE}{dx}$ signatures along the length of the track, for the z-axis and y-axis exposures respectively, are found to be consistent with this simulation. 

\section{$^{252}$Cf Nuclear Recoils in a CYGNUS-m$^3$ Scale Vessel}

\begin{comment}

Following the successful characterisation of the MMThGEM-Micromegas detector in the test vessel and preliminary receptivity to a $^{252}$Cf neutron source, the detector assembly was transferred to the Kobe University C/N-1.0 vessel to perform an initial large scale demonstration in low pressure SF$_6$. In this section, the detector installation in the C/N-1.0 vessel is discussed and results of an exposure to a $^{252}$Cf neutron source (0.3 MBq), including supplementary SRIM and SREM (Stopping and Range of Electrons in Matter) \cite{SREM} simulations, are presented.

The C/N-1.0 vessel is a large CYGNUS-m$^3$ scale vessel which is capable of housing up to 18 small scale R\&D detector readout planes. An image of the C/N-1.0 vessel can be seen on the left of \autoref{fig:bento}. The vessel has a height and width of 1.6 m and features a central cathode design with a 0.5 m drift length either side. This large drift distance is conducive for testing different readout technologies with a full scale CYGNUS drift length. To this effect, small scale detector modules can be mounted to the nine panels, seen in \autoref{fig:bento}, either side of the cathode. The door of the test vessel is conveniently designed to fit into these panels, allowing the MMThGEM-Micromegas detector to be easily transferred to the central panel on one side of the C/N-1.0 vessel.

\end{comment}

\begin{figure}[h!]
    \centering
    \begin{subfigure}[c]{0.4\textwidth}
        \centering
        \includegraphics[trim={0 1.2cm 0 1.2cm},clip,width=\textwidth]{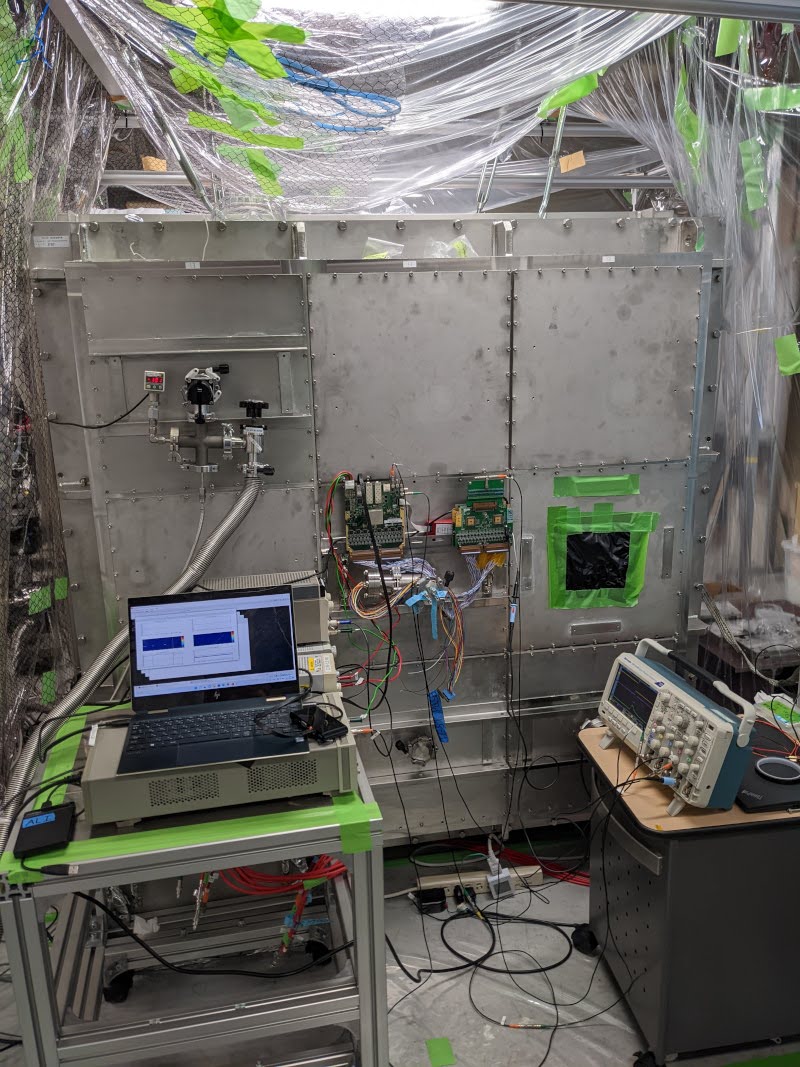}
        \label{fig:bentoimage}
    \end{subfigure}
    \hfill
    \begin{subfigure}[c]{0.5\textwidth}
        \centering
        \includegraphics[trim={0cm 0.75cm 0cm 1.3cm},clip,width=\textwidth]{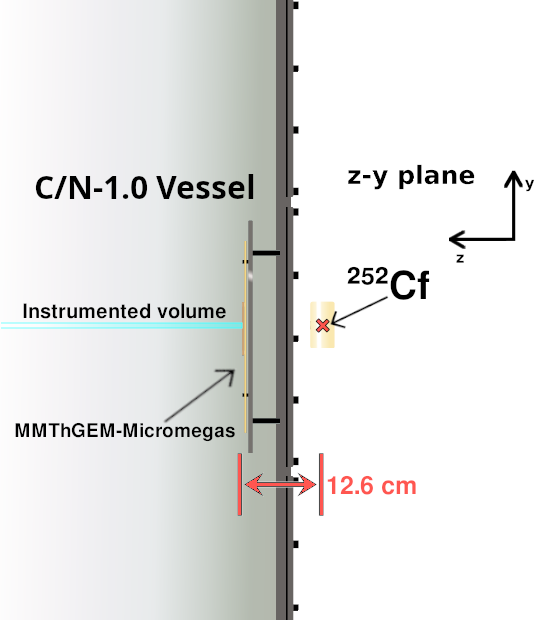}
        \label{fig:bentodiag}
    \end{subfigure}
    \caption{Image of the C/N-1.0 vessel following installation of the detector panel (left). Cross sectional diagram depicting the $^{252}$Cf neutron source position relative to the MMThGEM-Micromegas assembly in the C/N-1.0 vessel (right).}
    \label{fig:bento}
\end{figure}

Following successful characterisation, the MMThGEM-Micromegas detector was transferred from the test vessel to the Kobe University C/N-1.0 vessel to enable the measurement of NRs in a full-scale detector volume of low-pressure SF$_6$. The C/N-1.0 vessel, shown on the left of \autoref{fig:bento}, is a CYGNUS-m$^3$ scale vessel \cite{Vahsen2020} with back-to-back 1.6 m x 1.6 m x 0.5 m volumes. It can accommodate up to 18 small R\&D readout planes mounted on nine panels per side. The MMThGEM-Micromegas was installed centrally via the panel-compatible vessel door assembly. 

\begin{comment}
\begin{figure}[t!]

    \begin{subfigure}[b]{0.49\textwidth}
        \centering
        \includegraphics[trim={0 1.2cm 0 1.2cm},clip,width=\textwidth]{Images/BENTO_Vessel.jpg}
        \label{fig:bentoimage}
    \end{subfigure}
    \begin{subfigure}[b]{0.49\textwidth}
        \centering
        \includegraphics[trim={0cm 0.75cm 0cm 1.3cm},clip,width=\textwidth]{Images/BENTOSourceDiag.png}
        \label{fig:bentodiag}
    \end{subfigure}
    \caption{Diagram of the C/N-1.0 vessel depicting the vessel dimensions and installation of the MMThGEM-Micromegas detector (left). Image of the C/N-1.0 vessel following installation of the detector panel (right).}\label{fig:bento}
\end{figure}
\end{comment}

After installation, the C/N-1.0 vessel was evacuated and filled with 40 Torr of SF$_6$; this marks the first instance in which such a large volume of low pressure SF$_6$ has been utilised. Once filled, the field strengths were set to the optimised high gain settings of Configuration A (see \autoref{tab:field_configurations}) to observe the low energy NRs expected form the $^{252}$Cf source and take advantage of the gain calibration conducted in \autoref{sec:GasGain}. The $^{252}$Cf source was then placed externally in front of the central panel with a z-axis offset of 12.6 cm from the instrumented TPC volume, as illustrated on the right of \autoref{fig:bento}.

\begin{comment}
\begin{figure}[h!]
    \centering
    \includegraphics[trim={0cm 0.75cm 0cm 1.3cm},clip,width=0.5\textwidth]{Images/BENTOSourceDiag.png}
    \caption{Cross sectional diagram depicting the $^{252}$Cf neutron source position relative to the MMThGEM-Micromegas assembly in the C/N-1.0 vessel.}
    \label{fig:BENTOSourcePos}
\end{figure}
\end{comment}

\begin{figure}[b]
    \begin{center}
        \begin{subfigure}{0.49\textwidth}
            \centering
            \includegraphics[trim={0cm 0cm 0cm 2cm},clip,width=\textwidth]{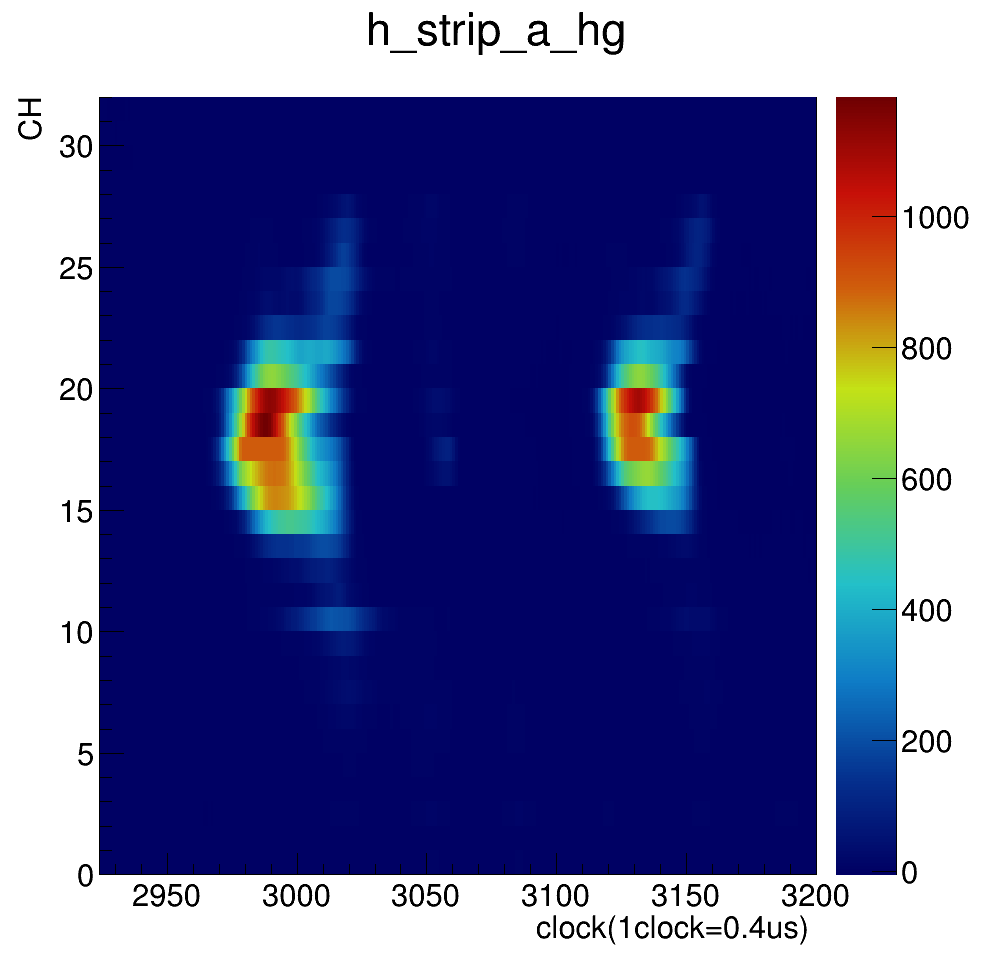}
        \end{subfigure}
        \begin{subfigure}{0.49\textwidth}
            \centering
            \includegraphics[trim={0cm 0cm 0cm 2cm},clip,width=\textwidth]{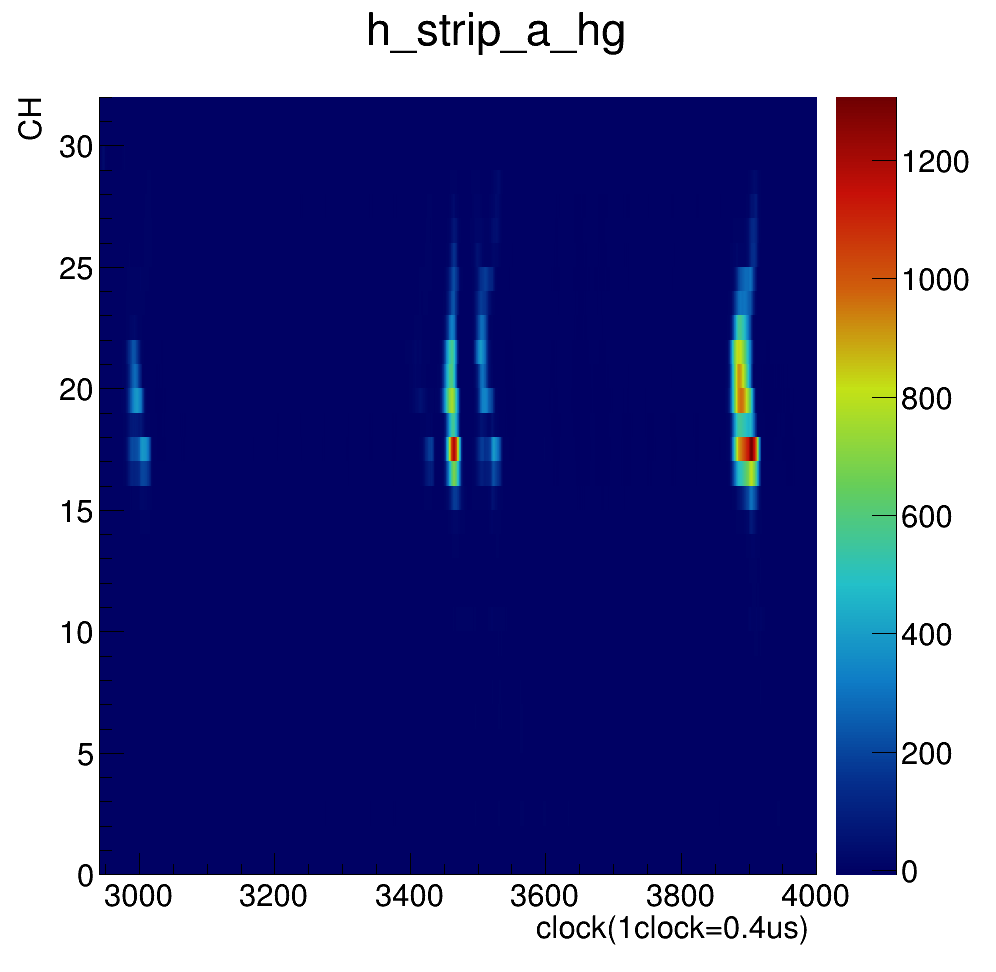}
        \end{subfigure}
        \caption{Examples of events which were captured during the $^{252}$Cf neutron source exposure in the C/N-1.0 vessel.}
        \label{fig:NR_Event_Inspection_Kobe}
    \end{center}
    \end{figure}

Events were acquired over the course of 13 hours and 1210 events were successfully captured. Two notable example events from the exposure can be seen in \autoref{fig:NR_Event_Inspection_Kobe}. The successful observation of events in such a large volume of low pressure SF$_6$ is a significant step forward for demonstrating the MMThGEM-Micromegas detector as a scalable readout technology for a future CYGNUS search. 

The event on the left  of \autoref{fig:NR_Event_Inspection_Kobe} consists of two symmetric clusters of charge, likely caused by the charge discontinuity effect discussed previously in \autoref{sec:Directional}. Interestingly, this particular event exhibits some preliminary evidence of the head-tail effect, with neutrons incident from the left-hand side. The event on the right of \autoref{fig:NR_Event_Inspection_Kobe} contains four charge clusters. This could be caused purely by the charge discontinuity effect, or it could be an indication of two SF$_5^-$ minority peaks. In the latter case, the charge in the smaller leading clusters contain $\sim$ 30\% of that in the larger respective trailing clusters, much larger than those seen previously at lower gas gains \cite{Phan2017}. This could suggest that the large gain of this detector constitutes a significant sensitivity increase to measurable minority peaks. These results are greatly promising for this detector and further work is required to investigate sensitivity to the head-tail effect and minority peaks in the future.

\begin{figure}[h!]
    \centering
    \includegraphics[trim={2cm 0.4cm 2cm 1.6cm},clip,width=0.95\textwidth]{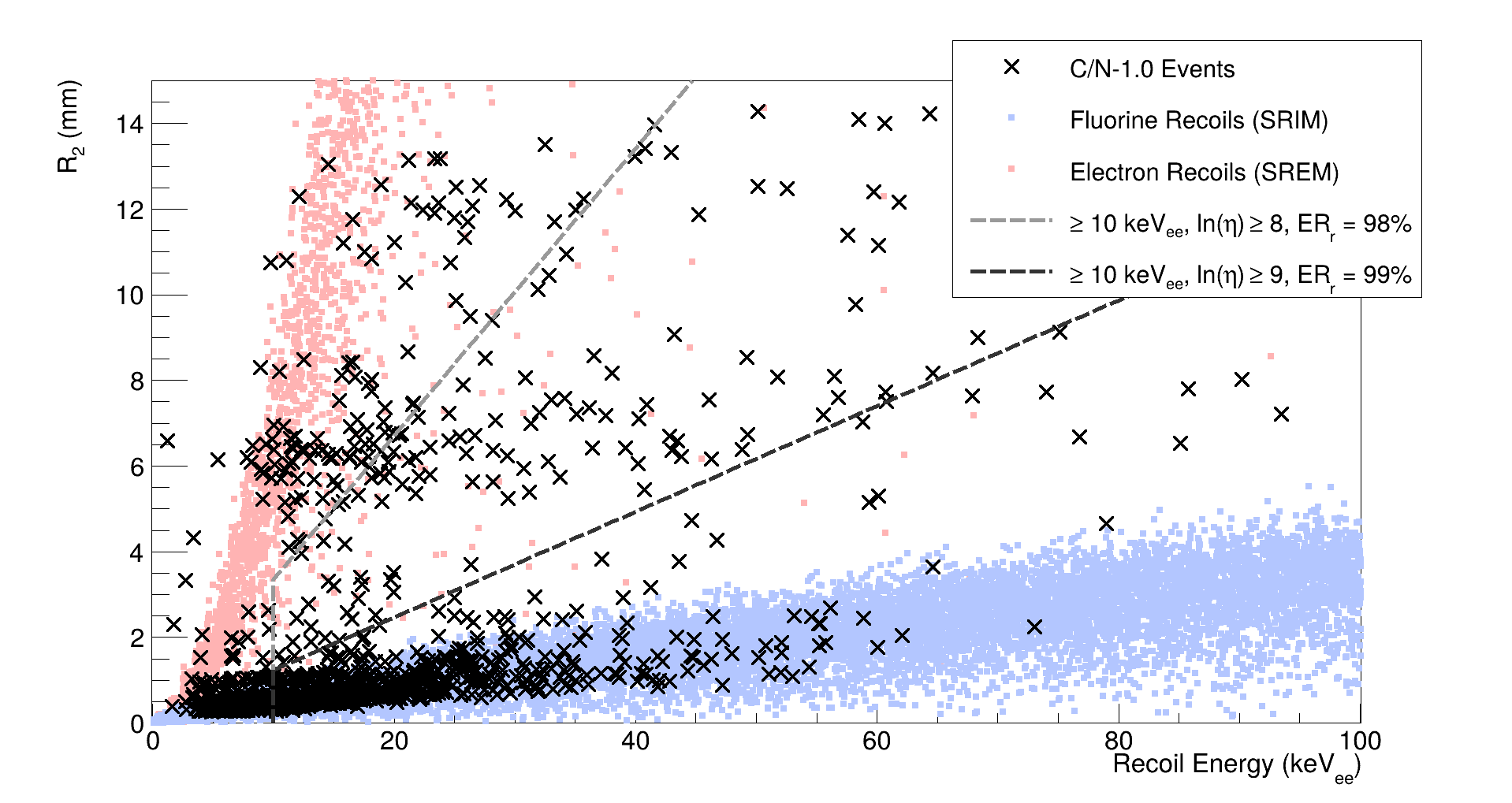}
    \caption{Two-dimensional range vs estimated recoil energy for events observed in the C/N-1.0 vessel during the $^{252}$Cf neutron exposure (black cross markers). The simulated NR and ER events from SRIM (pale blue) and SREM (pale red) are also observed along with the selection cuts described in the previous subsection.}
    \label{fig:BENTOranges}
\end{figure}

The recoil energies of measured events during the neutron run in the C/N-1.0 vessel were calculated by first converting the signal integral into charge, via the calibration, and then relating this measured charge to the initial amount of ionisation charge via the gas gain, determined in \autoref{sec:GasGain}. Finally this was related to the amount of electron equivalent energy via the SF$_6$ W-value. The 2-dimensional recoil range was then calculated by first determining the z-range from the points above a 40 mV threshold. The y-range was then determined via the TLR algorithm, described in \autoref{sec:Directional}, to minimise the influence of charge dissipation in the resistive layer. The resulting 2-dimensional range of each event can be seen plotted against the estimated recoil energy in \autoref{fig:BENTOranges}; indicated by the black cross markers.

Supplementary SRIM and SREM simulations of NRs and ERs respectively are also shown in \autoref{fig:BENTOranges}. Given the relative abundance of fluorine in SF$_6$, the recoil of 100 fluorine nuclei were simulated for every 1 keV increment between 1 and 150 keV in 40 Torr of SF$_6$ (pale blue); the energy was converted to keV$_{ee}$ via the Lindhard model \cite{Lindhard, Sorensen2015}. SREM was then used in a similar fashion to simulate 100 ERs for each 1 keV increment between 1 and 100 keV (pale red). A parameter was defined to facilitate a $\frac{dE}{dx}$ cut such that:

\begin{equation*}
    \eta = E/R_2, 
    \end{equation*}where \textit{E} is the electron equivalent energy in keV$_{ee}$ and $R_2$ is the 2-dimensional range in the z-y plane in metres. This is analogous to a parameter whose natural logarithm has previously been used for ER/NR discrimination in low pressure CF$_4$ \cite{Phan2016} and SF$_6$ \cite{LaflerThesis}. Using the simulated data, a strict cut which included events above 10 keV$_{ee}$ with $ln(\eta) \geq $ 9 and a more lenient cut with $ln(\eta) \geq $ 8 were made; with ER rejection of 99\% and 98\% respectively.

\begin{comment}

\begin{figure}[b!]
    \centering
    \includegraphics[trim={0cm 0cm 0cm 0cm},clip,width=0.7\textwidth]{Images/lnetacomparison_thesis_labelled3.png}
    \caption{ln($\eta$) plotted as a function of electron equivalent recoil energy for simulated fluorine NRs (blue) and ERs (red) in 40 Torr of SF$_6$. Cuts which are able to successfully reject 99\% (solid black line) and 98\% (dashed grey line) of ERs are superimposed on the simulated data.}
    \label{fig:simulationETA}
\end{figure}

\end{comment}

\begin{comment}
The natural logarithm of $\eta$ was calculated for each simulated event and plotted as a function of recoil energy, shown in \autoref{fig:simulationETA}. As can be seen, the populations of NRs and ERs are well separated by this parameterisation. In an attempt to quantify the ER rejection (ER$_r$), two cuts were imposed in this parameter space. Firstly, a strict cut which included events above 10 keV$_{ee}$ with $ln(\eta) \geq $ 9 (solid black line), and secondly a more lenient cut with $ln(\eta) \geq $ 8 (dashed grey line). It was found that the strict cut was capable of ER$_r$ to 99\% while the more lenient cut resulted in ER$_r$ to 98\% for the simulated events. %In the following subsection, these simulated recoils are compared to the recoils obtained during the $^{252}$Cf neutron exposure in the C/N-1.0 vessel.
\end{comment}

As shown in \autoref{fig:BENTOranges}, a significant portion of C/N-1.0 events appear to strongly correlate with the simulated NR band and fall well within the ER$_r$ = 99\% selection cut, indicating that a large portion of events are consistent with NRs. A number of events can also be seen to cluster close to the simulated ER band and fall outside the ER$_r$ = 98\% selection cut, suggesting that these events are consistent with ERs. Events can also be seen to fall between the two selection cuts and are neither consistent with the NR or ER band. Some of these could be a consequence of mishandling events which contain minority peaks or are not constrained to the instrumented area as well as diffusion and charge dissipation effects. Therefore, further work is required to investigate these events. This would include scaling up the number of instrumented channels in both x and y-strip planes for complete 3-dimensional event reconstruction, as well as exposure to a high energy gamma-ray source for a more explicit ER/NR discrimination study.

\begin{comment}
A significant portion of the C/N-1.0 events can be seen to fall within the more lenient, ER$_r$ = 98\%, simulated selection cut. It is noted that, events which occur closer to the cathode will experience more notable diffusion, artificially broadening the measured range of these events. As diffusion has not been accounted for here, some NRs could be found to fall within the more lenient simulated selection cut; although this diffusion is anticipated to be $\lesssim$ 1 mm \cite{Phan2017}. An alternative effect which could contribute to artificial range broadening of events, is the possible presence of unidentified minority peaks. Given the estimated ER$_r$ from this selection cut, these events are less likely to be attributed to ERs induced by background sources or the gamma-rays produced by the $^{252}$Cf source.  However, due to their longer ranges, ERs which are not completely captured within the instrumented volume could also occupy this area of the parameter space. 

\end{comment}

\begin{figure}[h!]
    \begin{subfigure}{0.49\textwidth}
        \centering
        \includegraphics[trim={0.4cm 0.3cm 1.4cm 1cm},clip,width=\textwidth]{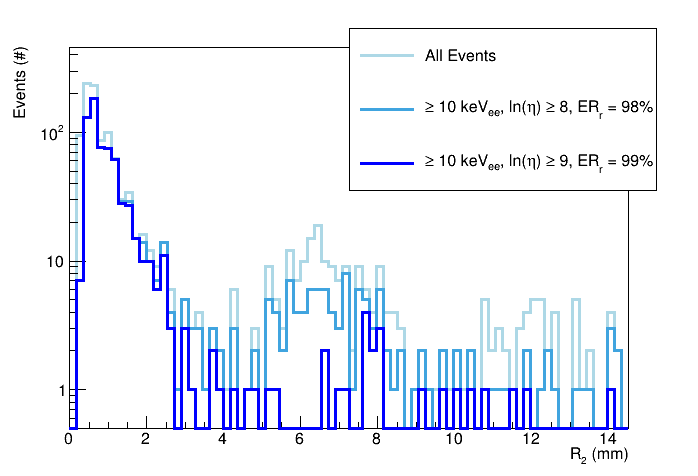}
        \label{fig:etaRangedist}
    \end{subfigure}
    \begin{subfigure}{0.49\textwidth}
        \centering
        \includegraphics[trim={0.4cm 0.3cm 1.4cm 1cm},clip,width=\textwidth]{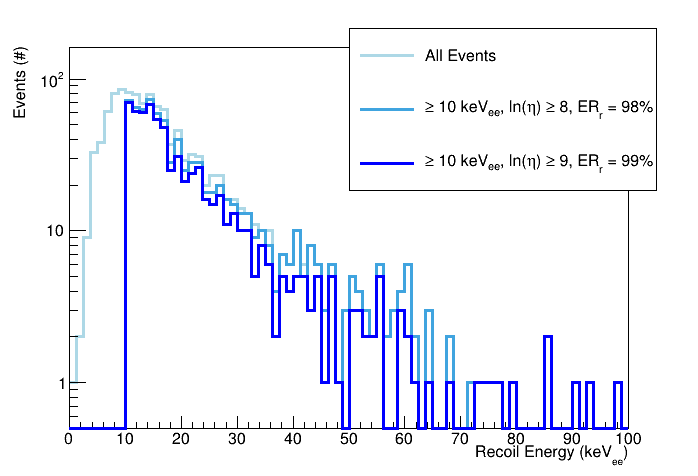}
        \label{fig:etaEnergydist}
    \end{subfigure}
    \caption{Histograms showing the measured ranges (left) and energies (right) of all recoil events (pale blue) and with the lenient (light blue) and strict (blue) simulated selection cuts applied.}
    \label{fig:KOBErangeandenergyHist}
\end{figure}

\autoref{fig:KOBErangeandenergyHist} presents a more focused view of the distribution of the measured recoil ranges (left) and energies (right) from the C/N-1.0 vessel. With no selection cuts applied, events appear to accumulate in three distinct clusters in the range distribution. Upon inspection, this was found to be a quantisation artefact induced by the coarse MMThGEM hole pitch, discussed in \autoref{sec:Directional}. This structure is not present following the strict selection cut, most likely due to the short (< 4 mm) range of NRs in this energy range. Finally, it is noted that the measured energy spectrum is consistent with that of a $^{252}$Cf neutron source demonstrated elsewhere in similar low pressure gaseous detector exposures \cite{Phan2016,Callum_thesis}. 
These findings provide strong evidence that NRs were successfully observed during operation in the large C/N-1.0 vessel. This result therefore represents a significant step towards realising this detector technology as a scalable readout option for a CYGNUS search.

%\autoref{fig:KOBErangeandenergyHist} presents a more focused view of the distribution of the measured recoil ranges (left) and energies (right) from the exposure in the C/N-1.0 vessel. Interestingly, with no selection cuts applied, events appear to accumulate in three distinct clusters in the range distribution. Upon inspection, it was observed that the majority of events with ranges less than 4 mm contained 1-2 clusters of arriving charge, while those between 4 and 10 mm exhibited 2-3 clusters, and ranges exceeding 10 mm featured 3 or more clusters of arriving charge. This result indicates that the discontinuity caused by the MMThGEM hole pitch, discussed in \autoref{sec:Directional}, is likely inducing a range quantisation artefact. This structure is not present following the strict selection cut, most likely due to the short (< 4 mm) range of NRs in this energy range. Finally, it is noted that the measured energy spectrum is consistent with that of a $^{252}$Cf neutron source demonstrated elsewhere in similar low pressure gaseous detector exposures \cite{Phan2016,Callum_thesis}. The combination of the range quantisation, short lengths of NRs, and the shape of the $^{252}$Cf energy spectrum explains the large population of events with low energy and short ranges.

%These findings provide strong evidence that NRs were successfully observed during operation in the large C/N-1.0 vessel. This result therefore represents a significant step towards realising this detector technology as a scalable readout option for a CYGNUS search. In the future, the confirmation of this finding would benefit from the instrumentation of more channels, including both x and y-strip planes, and an additional exposure to a high energy gamma-ray source for the purpose of an explicit ER/NR discrimination study. 

\section{Conclusions}
\label{sec:Conclusions}

In conclusion, the operation of a coupled MMThGEM-Micromegas detector in low pressure SF$_6$ was discussed in this paper. 32 y-strips were individually instrumented, demonstrating the largest area of individually instrumented strips in an NID gas with this detector to date. This was made possible by the large gas gain of 1.22 $\pm$ 0.08 $\times$ 10$^5$. This is an improvement on the previous measurements with the isolated MMThGEM and demonstrates the first charge amplification of order 10$^5$ in an NID gas; significantly larger than typical order 10$^3$ NID gas gains previously seen. Complete 2-dimensional directionality was also demonstrated with alpha particle tracks by reconstruction of the principal axis, through development of a total linear regression algorithm; and by sense recognition, where $\frac{dE}{dx}$ signatures were found to be consistent with simulated alpha tracks.
Following characterisation in the test vessel, the detector was installed in the CYGNUS-m$^3$ scale C/N-1.0 vessel and exposed to a $^{252}$Cf neutron source, with most observed events found to be consistent with simulated fluorine nuclear recoils. While further work is required for explicit ER/NR discrimination, head-tail sensitivity, and minority peak observations, these results represent an important step toward scaling this technology for a CYGNUS search. It is recommended for future designs to reduce MMThGEM hole pitch and remove the Micromegas resistive layer to better match the strip pitch of the Micromegas and mitigate charge dissipation effects.

%Following characterisation in the test vessel, the detector was then installed in the C/N-1.0 vessel, a large CYGNUS-m$^3$ volume, and exposed to a $^{252}$Cf neutron source. Through comparison to simulated fluorine NRs and ERs, the vast majority of events measured during this exposure were found to be consistent with simulated NRs; although, further work is required to explicitly determine the ER/NR discrimination capability of this detector and investigate the detector's sensitivity to the head-tail effect, for directional sense, and the possible observation of minority peaks, for fiducialisation. These findings demonstrate a significant step towards scaling up this readout technology for operation in a CYGNUS search. Finally, it is recommended that future design iterations of the coupled detector focus on the reduction of the MMThGEM hole pitch/diameter and removal of the resistive layer of the Micromegas. This is due to challenges associated with track discontinuity, charge dissipation, and range quantisation. 

\acknowledgments
The authors would like to acknowledge the "University of Sheffield EPSRC Doctoral Training Partnership (DTP) Case Conversion Scholarship" awarded to A.G. McLean. This work was also partially supported by the Japanese Ministry of Education, Culture, Sports, Science and Technology, Grant-in-Aid (24K07061 and 25K01025).
%\paragraph{Note added.} This is also a good position for notes added
%after the paper has been written.


%\bibliographystyle{unsrt}
%\bibliography{reference}

% MANUAL REFERENCING
%\input{REFERENCES}

% Please avoid comments such as "For a review'', "For some examples",
% "and references therein" or move them in the text. In general,
% please leave only references in the bibliography and move all
% accessory text in footnotes.

% Also, please have only one work for each \bibitem.

\end{document}